\def\be{\begin{equation}}
\def\ee{\end{equation}}
\documentstyle[aps,epsf,prl]{revtex}

\begin{document}
\twocolumn
[
\draft
\title{Self-Organized Critical Directed Percolation}
\author{Sergei Maslov$^{(*,\dag)}$ and Yi-Cheng Zhang$^{(*)}$}
\address{
(*) Institut de Physique Th\'{e}orique,
Universit\'{e} de Fribourg, CH-1700, Switzerland \protect\\
($\dag$) Department of Physics, Brookhaven National
Laboratory, Upton, New York 11973\\
and Department of Physics, SUNY at Stony Brook, Stony Brook, New York 11974}
\date{\today}
\maketitle

\widetext
\advance\leftskip by 57pt
\advance\rightskip by 57pt

\begin{abstract}

We introduce and study a dynamic transport model

exhibiting Self-Organized Criticality. The novel concepts of
our model are the probabilistic propagation of activity
and unbiased random repartition of energy among the active
site and its nearest neighbors. For space dimensionality 
$d\geq 2$ we argue that the model
is related to $d+1$ dimensional directed percolation,
with time interpreted as the preferred direction.
\end{abstract}
]
\narrowtext

Directed Percolation (DP) is one of the simplest and most recurrent models
in statistical mechanics. Under very general guidelines (locality, scalar 
variable, etc.) Grassberger \cite{GRA}
and  Janssen \cite{JAN} have proposed
that a wide range of models would fall into the DP universality class. This
conjecture has stood the test of time. An impressive parade of models under
various disguises turned out to belong to the same DP class (see \cite{G95} 
for a review). However, they all 
share one common feature: the activation probability $p$ has 
to be defined in advance, and when it is 
properly fine-tuned, a phase transition takes place. Our aim here is to design
a dynamical model in which critical directed percolation would occur via
self-organization.

Consider the following activation-transport problem. A conserved
quantity called energy $E$ can be stored at each site of a 
$d$-dimensional square lattice with open boundaries. An input
energy $\delta E<1$ is added to the fixed input site in the center of
the system and this site is declared
$active$. An active site can propagate activity under the following
rules: 1) Random repartition of energy among the active site and its
$2d$ neighbors i.e. $E_i \rightarrow x_i\sum_{j=1}^{2d+1} E_j, (i=1,...,2d+1)$
, where 
$x_i=r_i/ \sum_{j=1}^{2d+1}r_j$, and $r_i$ are uncorrelated
random numbers between 0 and 1; 
2) After repartition $each$ of the above $2d+1$ sites becomes 
active with the $probability$ given by its energy content. If $E_i>1$
activation happens with certainty. There is no spontaneous activation 
of sites not connected to the current active site.
The above process is repeated until no active sites are left
in the system and then a new input is introduced.
The above process strictly conserves energy except for the activation at 
the boundaries. Open boundaries allow the energy to escape.

It is easy to convince oneself that 
the system eventually organizes itself
into a critical stationary
state where the average stored energy $\bar{E}<1$.
In this state the energy input and output through 
the boundaries 
balance each other, on average. The avalanches in this critical
state have no characteristic size except the size
of the system. 
The above model is inspired by the celebrated  Bak-Tang-Wiesenfeld (BTW)
model of Self-Organized Criticality (SOC) \cite{BTW}.
In fact it is a modified version of the SOC energy model
introduced earlier \cite{zhang_en}.
The new feature here is that the activation probability is given by
the transported energy.

\begin{figure}
\epsfxsize 12cm
\centerline{\epsfbox{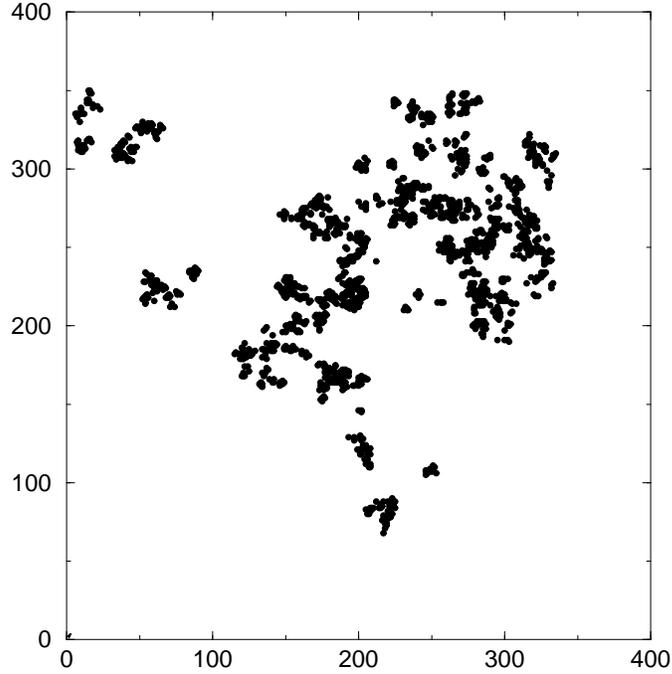}}
\caption{Snapshot of activity in one realization of the 
avalanche process in the system of size $400 \times 400$. 
Active sites follow a fractal statistics}
\label{fig_cl} 
\end{figure}

In our model there exists an effective probability
of activation of neighbors related to the
average energy $\bar{E}$, stored in the system.
Each of the $2d+1$ sites participating in a single
energy redistribution has equal chance to become active
at the next time step.
The propagation of activity in time
is reminiscent of the model of epidemic without
immunization in $d$-dimensions \cite{DP_disease},
which is a particular realization of $d+1$-dimensional 
bond directed percolation with the preferred direction 
interpreted as time. 
The stationary state of our model corresponds to the
critical disease propagation.
In Fig. \ref{fig_cl} we show a snapshot of the typical fractal pattern formed
by active sites at a given time step in the two dimensional version
of our model.
Below we shall demonstrate that despite all the above suggestive
considerations, the aim of mapping our model to Directed Percolation 
is only partially fulfilled: in $d\geq 2$ indeed 
the activation clusters seem to fall 
into the standard DP class; for the special case
$d=1$ however, though still critical, our model belongs to a  
new class, probably corresponding to DP with long range correlated disorder.

There are three quantities
characterizing an avalanche of activity: its spatial size $R$,
its temporal duration $T$, and its volume, equal to the total number of
activations $S$. They are connected via two scaling relations:
\begin{eqnarray}
S=R^D& \\ T=R^z& \qquad .
\end{eqnarray}
In the DP mapping these quantities are interpreted as 
the perpendicular and parallel size of cluster and its volume
correspondingly.
 
\begin{figure}
\epsfxsize 12cm
\centerline{\epsfbox{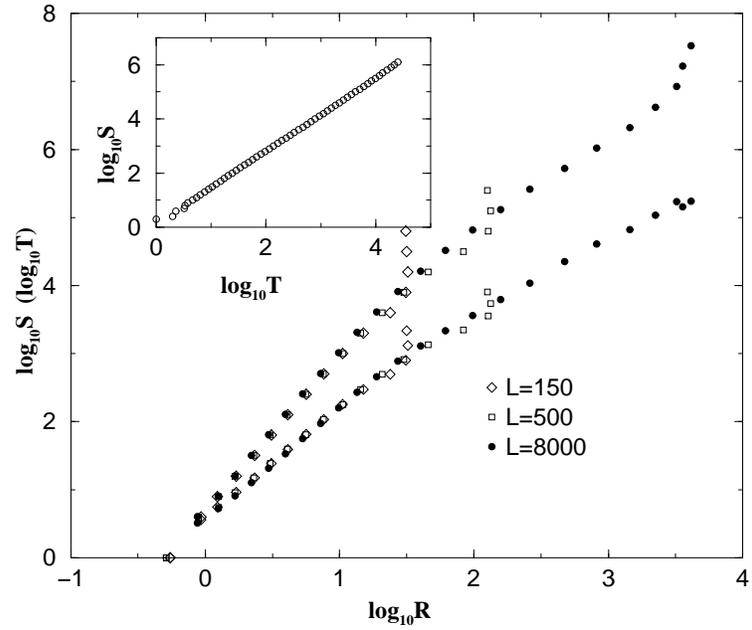}}
\caption{The scaling of $S$ and $T$ with $R$ in one dimension for
systems of different sizes. Crossover to  smaller values of exponents
$D$ and $z$ happens around $R=50$. The inset shows the log-log plot
of $S$ vs $T$ in the system of size $L=3000$. There is no crossover
observed and the scaling exponent $\alpha$ is measured to be $1.35
\pm 0.03$.}
\label{fig_1d} 
\end{figure}

We have measured $S$, $T$ and $R$ for the avalanches in our model.
The results of the numerical simulation in 1+1 dimension are shown
in Fig. \ref{fig_1d}. 
The exponent $D$ crosses over from 
$D=2.35 \pm 0.1$, which is close to the DP value ($2.32$ \cite{DP1}), 
to $1.3 \pm 0.1$. 
At the same time the exponent $z$ crosses over from $1.7 \pm 0.1$
to a much lower
value of $1.0 \pm 0.1$. It is interesting to note that 
$S$ vs $T$ plot has a fairly good scaling of the
form $S \sim T^{\alpha}$ throughout the whole region covered
by our simulations.  
The exponent $\alpha=1.35 \pm 0.03$,    
is different from its DP value $\alpha=D/z=1.47$ \cite{DP1}. This suggests
that in one dimension we are really dealing with a universality class
different from 1+1-dimensional DP.

\begin{figure}
\epsfxsize 12cm
\centerline{\epsfbox{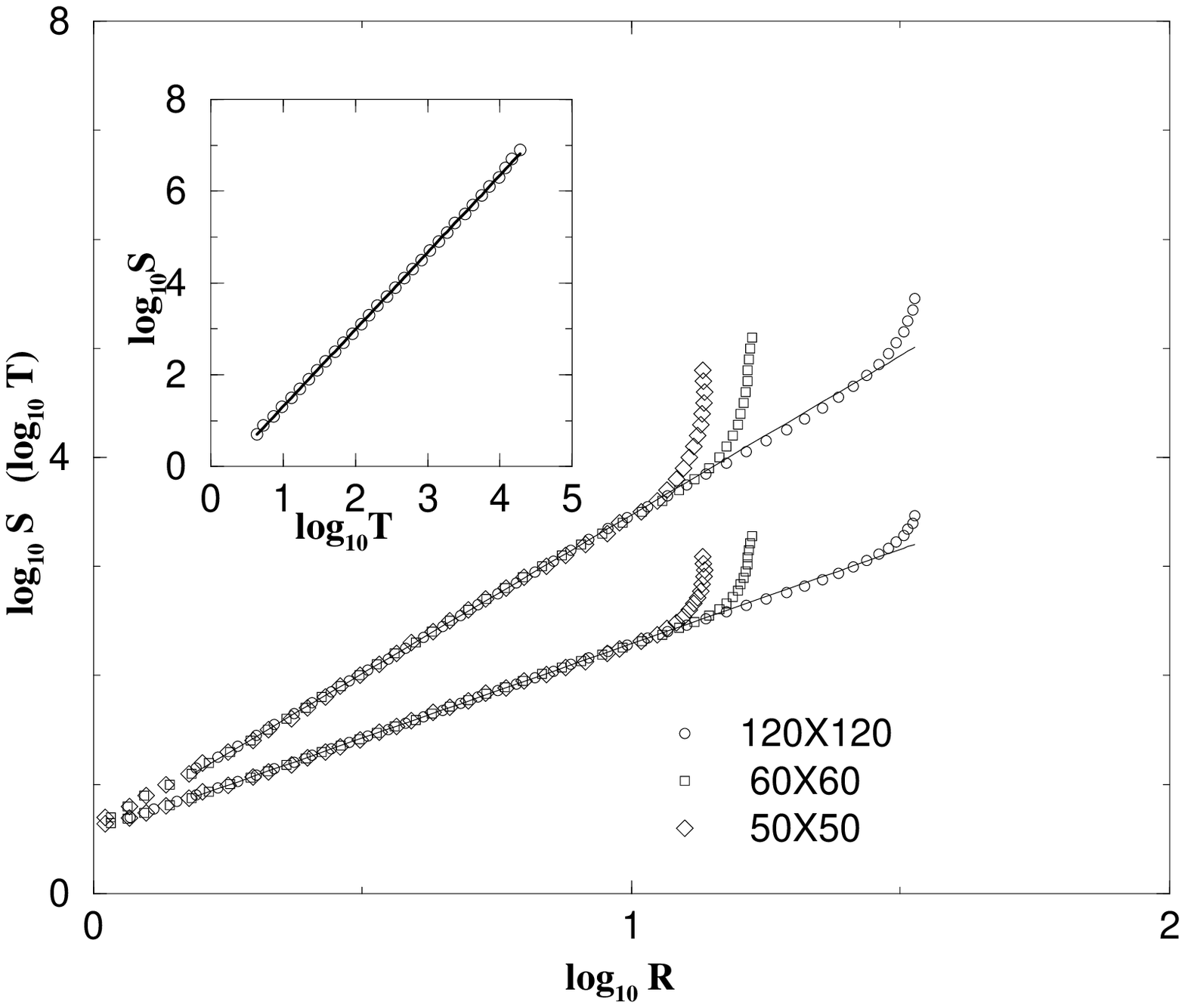}}
\caption{The scaling of $S$ and $T$ with $R$ in two dimensions for
systems of different sizes. Straight lines are fits, giving 
$D=2.92$ and $z=1.73$. The inset shows the scaling
of $S$ with $T$. The measured exponent is $\alpha=1.675 \pm 0.010$.}
\label{fig_2d} 
\end{figure}

In order to check further the relation between our model and DP
we have performed simulations in 2+1 dimensions.
The results are shown in Fig. \ref{fig_2d}.
Here we have measured
$D=2.92 \pm 0.05 $ and $z=1.73 \pm 0.05$ in fair agreement with the DP values
of $2.95$ and $1.76$ \cite{DP2}, respectively. No crossover was observed
for system sizes up to $120 \times 120$. The 
scaling between $S$ and $T$ is the cleanest and the measured
value of exponent $\alpha=1.675 \pm 0.01$ is in perfect agreement with its
DP value of $\alpha=D/z=1.675 \pm 0.01$ \cite{DP2}. 
We measured also the average stored energy $\bar {E}$ in critical state to be
$0.393 \pm 0.005$, for $d=1$, and $0.210 \pm 0.005$, for $d=2$. 

By definition, when the avalanche is over the energy at
every site in the system is smaller than unity. But while it is
running the energy is not bounded and in principle can
diverge.  
We expect that the larger is the avalanche size $S$ the larger 
is the maximal energy $E_{max}(S)$ reached during 
this avalanche. Our numerics indicate a very slow increase of
$E_{max}$ with $S$, for instance in 2+1 dimensions $E_{max}(S)$
increases from 1.18 to 1.75 while $S$ spans 6 decades.
The observed increase in $E_{max}(s)$ seems to be even slower than logarithmic,
we were not able to determine whether it would saturate or diverge slowly. 
In any 
case we can say that empirically events $E_i>1$ are rare and the values are 
reasonably bounded. 
The overwhelming majority of the sites have simple probabilistic
interpretation since $E_i\leq 1$.  

Another difference between our model
and DP is that while $p$ is a uniform
constant for DP models, in our model 
$E_i$ is space dependent. If a particular site or a group of sites contain
more energy than the average, 
this excess amount cannot immediately diffuse to infinity.
The neighborhood in question will be more prone to
activation, which may lead to spatial and temporal correlation in activity.
Indeed this is the case for $d=1$. 
To detect spatial correlation in 
$E_i$, we have studied the fluctuations of the convolution of the energy 
variable: 
$\sigma(x)=\sum_{i=1}^x (E_i-\bar{E})$. This method is known to
be able to detect the presence of very weak correlations in $E_i$ 
\cite{Zhang_fract}.
For uncorrelated energies we would have
$\sqrt{\langle \sigma(x)^2 \rangle} \sim x^{0.5}$.
Any deviation of the scaling exponent from $0.5$ will indicate the
presence of spatial correlation in $E_i$. 
In the one dimensional version of our model
we have found this exponent to be in the range
between $0.33$ and $0.40$, quite far from the uncorrelated value  
$0.5$. This implies that $\sigma(x)$ performs
so-called "persistent" random walk, being attracted to the regions it
has visited before \cite{Mandelbrot}.

At first sight our model should be compared to the DP model with quenched
disorder \cite{q_dp}, where the probability
of activation $p(i)$ is given for each site and never changes.
In our model it is true that $E_i$ does not update when the
activation is absent, thus in principle has a long memory. 
Activation reshuffles the energies
in the neighborhood and $E_i$ becomes "annealed" as soon as it
visits this site or its nearest neighbors.
The conserved nature however implies that
these are not completely fresh random
numbers, as a genuine DP model would require. 
For the model to belong to DP universality class the 
excess of energy should be able to diffuse away fast enough.
For $d=1$ as discussed above, this is not true and
the spatial correlation of $E_i$ is relevant. 
The excellent agreement of the exponents of our model in $2$ dimensions
with $2+1$-dimensional DP suggests that in high dimensions all
these correlations are irrelevant.
We conjecture that for $d\geq 2$
our model can effectively diffuse away any energy build-up and
hence it belongs to the universality class of DP.

On a coarse-grained level the energy is very uniformly distributed.
This homogeneity is remarkable especially taking into account
that the energy input in our simulations is
highly inhomogeneous: the energy is
injected only at the central site. We have checked that
the exponents do not change when the energy is added
to a randomly selected site each time, much
in the spirit of BTW model. 
This is another manifestation of the model's ability
to organize itself to the unique critical state, independent of initial
state or a particular type of driving.

In traditional SOC models activation occurs via a threshold mechanism.
Though it is possible that in $Nature$ threshold activation mechanisms play an
important role, the probabilistic activation rules are surely more realistic: 
A site containing less energy than the threshold
can still be activated by its active neighbors, but this
occurs less likely. Our model can be called the probabilistic SOC.
Probabilistic activation alone does not make any model
to fall into the DP universality class. 
It is important that we use the {\it neutral} redistribution
rule, when each of $2d+1$ sites involved
in redistribution gets an unbiased random share of the conserved energy.
In traditional SOC models, like
for instance in BTW-like models 
an active site always loses its energy to its
neighbors. Therefore, during one
avalanche its chances to be activated again are smaller than for 
its neighbors. We call such redistribution rules $charitable$.
This leads to an 
effective repulsion of currently active sites from the previously activated
region or outward bias in the spread of activity. Indeed, the
dynamic exponent $z \sim 1.33$ in 2-dimensional 
BTW model with the charitable rule of dynamics is smaller than 
our $z=1.73$. That means
that the outward expansion of activity is faster in BTW model due to repulsion.
It is interesting to mention that simple balance of the velocity of
moving active front to the total repulsion force acting on it gives the
expression $z_{BTW}={d+2 \over 3}$ \cite{zhang_en} which agrees very well 
with the results of numerical simulations of the BTW model. 
Finally a $greedy$ redistribution rule where currently
active sites on average gain energy can be envisaged.
It is anticipated \cite{zhang_91} that models with  
$charitable$, $neutral$ and $greedy$ rules should belong 
to different universality
classes. 
The detailed analysis of how the above three
different redistribution rules as well as threshold
or probabilistic activation influence the
universality class of SOC models will be relegated elsewhere.

In summary, we have introduced a simple probabilistic
SOC model which is a modified version of the energy model
\cite{zhang_en}. The energy is introduced
to the system at the central site, transported around 
through the conservative repartition of energy between 
currently active site and its nearest neighbors and dissipated at the 
boundary. 
The novel concepts of our model are: 
\begin{description}
\item[ 1)] A neutral rule
of repartition of energy, where each site involved in it gets an unbiased 
random share of the strictly conserved sum.
\item[ 2)] The probability of the site to be activated
by its active neighbor is given by its 
energy.
\end{description}
The system organizes itself to the critical state characterized by 
the power law distribution of avalanche sizes. 
The propagation of activity in time within the avalanche is conjectured 
to be equivalent to a model of spread of disease without immunization, which 
is a particular realization of bond directed percolation in $d+1$ 
dimension with preferred direction interpreted as time.  

The work of one of us (SM)
was supported by the U.S. Department of Energy Division
of Material Science, under contract DE-AC02-76CH00016. SM thanks
Institut de Phisique Th\'{e}orique, Universit\'{e} de Fribourg
for hospitality during the visit when this work was accomplished.
We thank Per Bak for useful comments and careful reading of
the manuscript.

\vfill\eject
\widetext
\onecolumn


\begin{references}

\bibitem{GRA} P.~Grassberger, Z.Phys. B {\bf 47}, 365 (1982).

\bibitem{JAN} H.~K.~Janssen, Z. Phys. B {\bf 42}, 151 (1981).

\bibitem{G95} P.~Grassberger, J. Stat. Phys. {\bf 79}, 13 (1995).

\bibitem{BTW}
P. Bak, C. Tang, and K. Wiesenfeld, Phys. Rev. Lett.
{\bf 59}, 381 (1987); \\ and Phys. Rev. A {\bf 38}, 364 (1988).

\bibitem{zhang_en}
Y.-C.~Zhang, Phys. Rev. Lett. {\bf 63}, 470 (1989).

\bibitem{DP_disease}
P.~Grassberger, Math. Biosci. {\bf 62}, 157 (1982); and 
J. Phys. A {\bf 17}, L215 (1985).

\bibitem{DP1}
J.W.~Essam, K.~De'Bell, J.~Adler and F.~Bhatti,
Phys. Rev. B {\bf 33}, 1982 (1986).

\bibitem{DP2}
P.~Grassberger, J.~Phys.~A {\bf 22}, 3673 (1989).

\bibitem{Zhang_fract} 
A.~Schenkel, J.~Zhang anf Y.-C.~Zhang, Fractals {\bf 1}, 41 (1993).

\bibitem{Mandelbrot}
B. B. Mandelbrot, {\it The Fractal Geometry of Nature}
(W.H. Freeman, San Francisco, 1982).

\bibitem{q_dp}
A.~J.~Noest, Phys. Rev. Lett. {\bf 57}, 90 (1986).

\bibitem{zhang_91}
L.~Pietronero, P.~Tartaglia and Y.-C.~Zhang, Physica A {\bf 173}, 22 (1991);\\ 
and Y.-C. Zhang, unpublished 1989.

\end{references}
\end{document}